# Analyzing the citation characteristics of books: edited books, book series and publisher types in the Book Citation Index

Daniel Torres-Salinas[1], Nicolás Robinson-García[2]*, Álvaro Cabezas-Clavijo[3], Evaristo Jiménez-Contreras[2]

[1] *torressalinas@gmail.com*
EC3: Evaluación de la Ciencia y de la Comunicación Científica, Centro de Investigación Biomédica Aplicada, Universidad de Navarra (Spain)

[2] *{elrobin, evaristo}@ugr.es, acabezasclavijo@gmail.com*[3]
EC3: Evaluación de laCiencia y de laComunicación Científica, Departamento de Información y Documentación, Universidad de Granada (Spain)

* To whom all correspondence should be addressed

**Abstract**
This paper presents a first approach to analyzing the factors that determine the citation characteristics of books. For this we use the Thomson Reuters' Book Citation Index, a novel multidisciplinary database launched in 2011 which offers bibliometric data on books. We analyze three possible factors which are considered to affect the citation impact of books: the presence of editors, the inclusion in series and the type of publisher. Also, we focus on highly cited books to see if these factors may affect them as well. We considered as highly cited books, those in the top 5% of those most highly cited in the database. We define these three aspects and present results for four major scientific areas in order to identify differences by area (Science, Engineering & Technology, Social Sciences and Arts & Humanities). Finally, we report differences for edited books and publisher type, however books included in series showed higher impact in two areas.

## 1. Introduction

One of the basic outcomes from the field of bibliometrics and citation analysis is the characterization of document types and field-based impact which allows fair comparisons and a better understanding of the citation patterns of researchers (Bar-Ilan, 2008). These studies are of great relevance in the field as they contextualize the impact of research as well as 'anomalies' such as the higher impact of reviews or editorial material versus research papers (Archambault & Larivière, 2009), the impact of research collaboration (Lambiotte & Panzarasa, 2009), or the importance of monographs in the Humanities (Hicks, 2004). In this context, the role played by citation indexes in general and by those developed by Eugene Garfield and constructed by Thomson Reuters in particular, have been of key importance in developing these analyses (Garfield, 2009). However, these citations indexes are mainly devoted to scientific journals and neglect other communication channels such as monographs. In this respect, Campanario et al. (2011) analyze the effect different document types have on Journal Impact Factors.

Many attempts have been made to analyze the impact of books using not only citation analysis but alternative metrics as well as databases. White et al. (2009) developed the concept of 'libcitation' in which they conceive library holdings as a measure of the impact of books and, hence, use this indicator to compare research units. Following this line of thought, Linmans (2010) suggests the use of library holding analysis to complement citation analysis when evaluating research units, especially in the Social Sciences and the Humanities. Torres-Salinas & Moed (2009) studied this idea in depth, exploring what they termed Library Catalog Analysis, which seeks to develop a bibliometric methodology based on the presence of books in online catalogues rather than in citations. Finally, Zuccala & van Leeuwen (2011) explored the potential of including book reviews in bibliometric research evaluations for the Humanities.

Regarding the use of citation analysis, Hammarfelt (2011) explores the possibility of using references from journal articles from the Web of Science to books, which are non-citable items, in order to analyze highly cited monographs. He highlights the difficulties this methodology entails when focusing on the impact of specific research units. Kousha, Thelwall & Rezaie (2011) go a step further and explore the use of Google Books and Google Scholar as alternative databases for retrieving, not only citations from journal articles to books, but also from books to books, showing a greater coverage than other databases such as Scopus.

Despite all these efforts, little is known of the characterization of book citation patterns. What is more, most of these studies are based on small data sets and specific disciplines. For instance, Cronin, Snyder & Atkins (1997) compared a list of highly influential authors derived from journal citations with another derived from books in the field of Sociology, concluding that these two publication types reflect complementary aspects of a fragmented picture. Tang (2008) took a step further and investigated the citation characteristics of a sample of 750 monographs in the fields of Religion, History, Psychology, Economics, Mathematics and Physics, finding significant differences when compared with findings in the literature on citation in journal articles. Georgas & Cullars (2005) adopted a different approach and analyzed the citation characteristics of literature in Linguistics to see if researchers' habits in this field were more closely related to the Social Sciences than the Humanities. In general, the conclusions of these studies must always be taken with caution as they cannot be extended to all scientific fields.

The launch in 2011 of the Thomson Reuters' Book Citation Index (henceforth BKCI) - which provides large sets of bibliometric data on monographs - was intended to resolve issues about data availability and may serve as a resource for further citation analyses and a more thorough understanding of the citation characteristics of books. It covers scientific literature since 1999 and, as with the other citation indexes, follows a rigorous selection process using the following principle criteria (Testa, 2010): currency of publications, complete bibliographic information for all cited references, and the implementation of a peer review process. Despite being very recent, the BKCI has already been the object of study. Indeed, many studies analyze its coverage, caveats and limitations. Leydesdorff & Felt (2012) analyze citation differences between monographs, edited volumes and book chapters by comparison with journal articles. They highlight the conceptual limitations that arise when book chapters are considered to be individual contributions and the manner in which this may affect any subsequent analysis of researcher output; the scarce number of citations that books receive; and the problems that arise from ignoring differences between book series and annual series.

The inflation of publication counts is reported by Gorraiz, Purnell & Glänzel (2013) who also point to limitations such as the lack of cumulative citation counts or the absence of affiliation data. Other studies of the BKCI are those reported by Torres-Salinas et al. (2012, 2013a, 2013b). Here a different perspective is taken, analyzing the database from the publishers' viewpoint: the field-specialization of publishers, concentration, citation patterns of book chapters and even suggesting the possibility of reproducing the Journal Citation Reports model by developing Book Publisher Citation Reports (Torres-Salinas et al., 2012). The present paper seeks to explore further aspects of books as a channel of scholarly communication and the citation characteristics that define them, by making use of the BKCI.

This study aims to compare the citation patterns of different book types. To do so, we define these in terms of three variables and present results for each of four major macro-areas of scientific knowledge (Science, Social Science, Engineering & Technology and Humanities). Specifically, our objective is to answer the following research questions:

1) Edited books vs. Non-edited books. There is a perception that edited books usually have a greater impact than non-edited books (Leydesdorff & Felt, 2012). In this study we aim to determine the extent to which this is true. Do differences exist between areas of knowledge?
2) Series books vs. Non-series books. The prestige or impact derived from the series in which the book is included could be considered evidence of the quality books may have due to the editorial process they have undergone in order to be published. Is there any empirical evidence for this claim in terms of higher citation rates?
3) Publisher type. Is the publishers' prestige related to a book's impact? Which publishers receive more citations: university presses, commercial publishers or other academic publishers?

Also, we take special interest in highly cited books. That is, the 5% of most cited books in each specific area. We try to determine the characteristics of this special subset of books in terms of our previously-defined variables.

## 2. Material and methods

This section is structured as follows. First, we describe the data retrieval and processing procedures, indicating the normalization problems encountered and how these were solved. In subsection 2.2, we define the areas under study and how these were constructed, basing our methodology on previous studies and offering an overview of the distribution of books by area in the BKCI as well as highlighting the main aspects of this database which must be taken into account when analyzing the results of our study. Then, in subsection 2.3, we define the variables analyzed and describe the methodology followed as well as the statistical analysis undertaken in order to pursue the goals of the study.

### 2.1. Data retrieval and processing, and definition of areas

Although this study is centered solely on books, due to the way the database is designed, we downloaded all records indexed as 'book chapter' and as 'book' in the BKCI. The reason for doing this is derived from the way the BKCI is designed and it means some modifications to the citation count are needed in order to overcome previously-mentioned limitations that we will now discuss. Book chapters are considered separate records; which means that publication counts are inflated, which benefits books with more chapters. Consequently, only records designated as books were included in the study. However, we cannot rely solely on the information provided by the records indexed as 'book' when considering citations. As mentioned above, citation counts are not cumulative so the citation data offered by the database for a given book does not contain those citations received for the book chapters. Hence, in order to obtain the total number of citations received by each book we must calculate the total number of citations received by its chapters.

The download took place in May 2012 and the study period was 2005-2011. The chosen time period is based on the availability of data at the time of retrieval. Data was entered into a purpose-built relational database. During data processing, publisher names were normalized as many had variants that differed depending on the location of their head offices in each country. For instance, Springer uses variants such as Springer-Verlag Wien, Springer-Verlag Tokyo, Springer Publishing Co, among others. Note that a fixed citation window was used, which means older books are more likely to be cited than others. Also, we emphasize the fact that citations included in the BKCI come from all the citation indexes provided by Thomson Reuters (SCI, SSCI and A&HCI) and not only the BKCI. Once the total number of book citations had been established we excluded Annual Reviews. This publisher, which includes a total of 234 records, was found to include journals rather than books, as noted by Torres-Salinas et al. (2013a) who observe that the citation pattern followed by this publisher is not comparable with that of the rest of the database. The final data set consisted of 28634 books.

*2.2. Overview and main limitations of the Book Citation Index*

In order to provide the reader with a general overview, we decided to cluster all BKCI subject categories (249) into four macro-areas: Arts & Humanities (HUM), Science (SCI), Social Sciences (SOC) and Engineering & Technology (ENG). Aggregating subject categories is a classical perspective followed in many bibliometric studies when adopting a macro-level approach (Moed, 2005; Leydesdorff & Rafols, 2009). These aggregations are needed in order to provide the reader with an overview of the whole database. Thus, we minimized the chances of overlapping for records assigned to more than one subject category. Also, we consider that these areas are easily identifiable by the reader as they establish an analogy with the other Thomson Reuters' citation indexes (Science Citation Index, Social Science Citation Index and Arts & Humanities Citation Index). With the exception of Sciences which, due to the heterogeneity of such a broad area, was divided into two: Science and Engineering & Technology. In table 1 we show the distribution of the sample of books analyzed across the four areas.

**Table 1. Distribution of books analyzed in this study by area as well as total and average citations received according to the Book Citation Index. 2005-2011.**

| Discipline | Acronym | Total Books | Total Chapters | Avg Chapters by book | %Books | Citations | Avg Citations |
|---|---|---|---|---|---|---|---|
| ENGINEERING & TECHNOLOGY | ENG | 3871 | 49076 | 12.68 | 14% | 34705 | 8.97 |
| ARTS & HUMANITIES | HUM | 8251 | 94825 | 11.49 | 29% | 52224 | 6.33 |
| SCIENCE | SCI | 9682 | 137027 | 14.15 | 34% | 241230 | 24.91 |
| SOCIAL SCIENCE | SOC | 10637 | 129754 | 12.20 | 37% | 99943 | 9.40 |
| **Total Books without duplicates** | | **28634** | **363152** | **13.25** | **100%** | **392429** | **13.70** |

Finally, we underline the main limitations of the BKCI. As a recent product, we consider it necessary to summarise the key aspects indicated in previous studies as the coverage and characteristics of this database may well influence the results derived from this study. Hence these aspects should be borne in mind when observing the results derived from our study. These can be summarized as follows:

- **Language bias**. The BKCI has a strong bias towards English speaking countries (96% of the records in the database are in English language) and specifically towards the United States and the United Kingdom as place of publication (these two countries represent 75% of the database). This seriously affects certain fields especially in the Social Sciences and the Humanities which are characterized by a strong national orientation (Hicks, 2004).

- **Great concentration of publishers.** The database contains records from up to 18 publishers, with a high presence of commercial publishers to the detriment of university presses and other academic publishers. Also, three publishers (Springer, Palgrave and Routledge) represent half of the database.

- **Inclusion of Annual Reviews as books.** The BKCI includes Annual Reviews with their ISSN considering them books. This generates a considerable distortion in terms of citations because reviews tend to accumulate a greater number of citations.

- **Inflation on publication counts.** As discussed in the Introduction, the BKCI considers book chapters and books to be independent records, meaning that publication counts may be misleading as this suggests that authors of books with a large number of chapters have a greater number of publications than those with fewer chapters (for further discussion of this issue the reader is referred to Leydesdorff & Felt, 2012).

- **Dispersion of citations**. Due to the distinction between books and book chapters, citations to each of them are also considered as independent. This means that citations received by book chapters are not included in the citation counts of books, therefore these must be added when working solely with books.

- **Citation errors.** Torres-Salinas et al. (2013b) and Gorraiz, Purnell & Glänzel (2013) reported errors in the citation count when comparing those included in the BKCI with those found using the Cited Reference Search option.

- **Conceptual problems.** Unlike journal articles, books present many issues which must be resolved before treating them for bibliometric purposes. For instance, should we consider new editions of a book as separate records? Or, must translations of a book be included in the same record? Depending on the solution reached this may lead to new technical problems such as assigning more than one publisher to a record (if translations are carried out by different publishers).

*2.3. Definition of variables and indicators*

Here we define and describe the three variables analyzed to characterize book citations: presence of editors, inclusion of books in a series and publisher type.

**Presence of editors.** In order to analyze edited and non-edited books we considered as the former those which were indexed as such according to the Book Editor (ED) field provided by the BKCI. We considered non-edited books those which had no information in this field. For instance, the book entitled 'Power Laws in the Information Production Process: Lotkaian Informetrics' which is single-authored by L. Egghe has no information in the ED field, therefore it is considered as a non-edited book. In contrast, the book 'Web 2.0 and Libraries: Impacts, Technologies and Trends' is edited by D. Parkes and G. Walton and has contributions from different authors, therefore it is considered an edited book.

**Inclusion in a series.** In order to analyze the inclusion of books in a series we used the field defined in the BKCI as Series (SE), tagging as such those records which contained information in this field and as non-series, those which did not. We identified a total of 3374 different series in the BKCI. The series with a higher number of books indexed in the BKCI for each field are: 'Studies in Computational Intelligence' published by Springer (243 books) for Engineering & Technology, 'New Middle Ages' by Palgrave (49 books) in Arts & Humanities; 'Methods in Molecular Biology' by Humana Press Inc (232 books) in Science, and 'Chandos Information Professional Series' by Chandos (118) in Social Sciences.

**Publisher type.** After data processing and normalizing publisher names as mentioned above, 280 publishers were identified in the BKCI. These publishers were then divided into the following three categories:

- University Press. Defined as any publisher belonging to a University such as the Imperial College Press or Duke University Press.

- Non-University Academic Publisher. Publishers belonging or related to organizations such as research institutions, scientific societies or any other type of entity not linked to universities such as the Royal Society of Chemistry or the Technical Research Centre Finland.

- Commercial Publisher. Publishers considered in this group are those not related to universities or any other scientific entity but to profit-oriented firms such as Routledge or Elsevier.

Finally, we characterized the factors that determine book citations. We report the effect size of citations for each variable and by area as a means of analyzing the extent of citation differences. For this, we use Cohen's *d* and *r*, and the criteria proposed by Cohen (1988). Hence, effect size is considered small, medium or large according to the following values:

$r$ effects: small $\geq 0.10$   medium $\geq 0.30$   large $\geq 0.50$

$d$ effects: small $\geq 0.20$   medium $\geq 0.50$   large $\geq 0.80$

Furthermore, we analyzed the characteristics of Highly Cited Books (henceforth HCB), that is, the 5% most highly cited for each of the four macro-areas under study. We identified 1534 books as HCB.

## 3. Results

In this section we report the results of the study according to the three variables analyzed. Firstly, we offer the results for presence of editors; secondly, we show those for the inclusion of books in series; and thirdly, we focus on publisher type. In the fourth subsection, we report the effect sizes for each variable. Finally we characterize the top 5% of HCB in terms of these variables.

### 3.1 Edited vs. Non-edited books

In table 2 we offer an overview of the sample of books analyzed according to the presence of editors. Overall, from the total sample (ALL), 12646 books (44%) had been edited while 15988 books (56%) had not. Edited books have a significantly higher citation rate than those which are non-edited, as shown by the average and median values. This occurs in the four areas studied. The most significant differences are found in the area of Science where edited books have an average of 35.41 citations per book as opposed to 10.16 citations per non-edited book. Also, edited books have a higher average of chapters per book and reach higher citation values as indicated by the standard deviation and median values. To a lesser extent, this also occurs in the Social Science and Engineering & Technology areas. The lowest differences between edited and non-edited books are found in the area of Arts & Humanities, where edited books have a citation average of 7.61, while non-edited books have an average of 5.81.

Table 2. Citation and statistical indicators. Edited vs. Non-edited books. 2005-2011

| Discipline | Type | Books | % Books | Avg Chapters by book | Citation Avg | Std Dev | Median |
|---|---|---|---|---|---|---|---|
| ALL | Edited Books | 12646 | 44% | 16.32 | 21.81 | ± 99.35 | 5.00 |
| ALL | Non Edited Books | 15988 | 56% | 9.81 | 7.16 | ± 7.61 | 2.00 |
| ENG | Edited Books | 1841 | 48% | 15.93 | 12.00 | ± 24.59 | 4.00 |
| ENG | Non Edited Books | 2030 | 52% | 9.71 | 6.21 | ± 15.82 | 1.00 |
| HUM | Edited Books | 2384 | 29% | 16.20 | 7.61 | ± 15.26 | 3.00 |
| HUM | Non Edited Books | 5867 | 71% | 9.59 | 5.81 | ± 14.45 | 2.00 |
| SCI | Edited Books | 5658 | 58% | 16.97 | 35.41 | ± 145.45 | 7.00 |
| SCI | Non Edited Books | 4024 | 42% | 10.30 | 10.16 | ± 35.96 | 2.00 |
| SOC | Edited Books | 4254 | 40% | 15.56 | 12.0 | ± 29.24 | 4.00 |
| SOC | Non Edited Books | 6383 | 60% | 9.96 | 7.66 | ± 24.35 | 2.00 |

*3.2 Inclusion in series vs. non-inclusion in series*

There are a total of 17789 books included in series (62% of the total share) while those not included in series amount to 10845 (38%) (Table 3). The distribution of books in series varies according to the area. Science and Engineering & Technology are the areas with the highest shares, especially the latter where books in series represent 71%. With regard to citation average and median values of books included in series, these two areas - and especially Science - show the most significant differences. In contrast, there are no significant differences in the Social Sciences, and the median value for both included and non-included books, is 3.00. The only exception noted is in Arts & Humanities, where non-included books have a higher citation average and median value than those included in series. Regarding the average number of book chapters by book, books not included in series have a higher average number of chapters, however, when observing each area we see that this is the case only for Science, whereas there are almost no differences in the other areas.

**Table 3. Citation and statistical indicators. Included in series vs. Non-included in series books. 2005-2011**

| Discipline | Type | Books | % Books | Avg Chapters by book | Citation Avg | Std Dev | Median |
|---|---|---|---|---|---|---|---|
| ALL | Series Books | 17789 | 62% | 13.41 | 12.62 | ± 45.68 | 3.00 |
| ALL | Non Series Books | 10845 | 38% | 15.88 | 10.98 | ± 95.38 | 3.00 |
| ENG | Series Books | 2746 | 71% | 12.59 | 10.06 | ± 28.25 | 3.00 |
| ENG | Non Series Books | 1125 | 29% | 12.86 | 7.6 | ± 23.50 | 2.00 |
| HUM | Series Books | 4585 | 56% | 11.24 | 5.91 | ± 14.43 | 2.00 |
| HUM | Non Series Books | 3666 | 44% | 11.83 | 6.86 | ± 15.04 | 3.00 |
| SCI | Series Books | 6349 | 66% | 12.92 | 29.63 | ± 69.19 | 5.00 |
| SCI | Non Series Books | 3333 | 34% | 16.54 | 15.93 | ± 169.37 | 2.00 |
| SOC | Series Books | 5854 | 55% | 12.03 | 9.1 | ± 27.10 | 3.00 |
| SOC | Non Series Books | 4783 | 45% | 12.40 | 9.75 | ± 25.76 | 3.00 |

*3.3 Type of publisher*

Overall, 83% of the books included in the BKCI belong to commercial publishers, followed at a considerable distance by university presses (14%) and non-university academic publishers (3%) (Table 4). This distribution varies substantially depending on the area. In Engineering & Technology the presence of commercial publishers is even higher (97%), while in the Arts & Humanities and the Social Sciences, the university presses have a higher presence (27% and 15%, respectively).

**Table 4. Citation and statistical indicators. Included in series vs. Non-included in series books. 2005-2011**

| Discipline | Type | Books | % Books | Avg Chapters by Book | Citation Avg | Std Dev | Median |
|---|---|---|---|---|---|---|---|
| ALL | Academic Non Univ | 919 | 3% | 13.60 | 23.90 | 53.40 | 5.00 |
| ALL | Commercial Publisher | 23843 | 83% | 14.51 | 12.36 | 39.67 | 2.00 |
| ALL | University Press | 3872 | 14% | 13.51 | 20.22 | 156.60 | 7.00 |
| ENG | Academic Non Univ | 72 | 2% | 14.67 | 16.22 | ± 28.45 | 5.00 |
| ENG | Commercial Publisher | 3726 | 97% | 12.61 | 8.62 | ± 18.90 | 2.00 |
| ENG | University Press | 57 | 1% | 13.84 | 23.96 | ± 65.73 | 7.00 |
| HUM | Academic Non Univ | 51 | 1% | 13.47 | 5.33 | ± 9.55 | 2.00 |
| HUM | Commercial Publisher | 5906 | 71% | 11.70 | 4.32 | ± 9.86 | 2.00 |
| HUM | University Press | 2270 | 28% | 10.92 | 11.62 | ± 22.20 | 6.00 |
| SCI | Academic Non Univ | 696 | 7% | 13.51 | 28.26 | ± 59.75 | 6.00 |
| SCI | Commercial Publisher | 8375 | 87% | 14.16 | 22.81 | ± 61.23 | 3.00 |
| SCI | University Press | 517 | 6% | 15.35 | 50.88 | ± 421.40 | 9.00 |
| SOC | Academic Non Univ | 181 | 2% | 7.44 | 10.71 | ± 24.24 | 3.00 |
| SOC | Commercial Publisher | 8816 | 83% | 12.43 | 7.33 | ± 21.13 | 2.00 |
| SOC | University Press | 1626 | 15% | 11.50 | 20.40 | ± 44.17 | 8.00 |

When analyzing citation average and median values in general, we find a pattern common to all areas: the commercial publishers are those with the lowest citation averages and the university presses are those with the highest number of citations. The highest difference is noted in the Arts & Humanities, where the latter show a citation average of 11.62 while the former have values of 4.32. The same is true of the Social Sciences, where university presses have a citation average of 20.40 versus commercial publishers, with 7.33. These differences are also significant for Engineering & Technology and Science, although not to the same extent. We do not observe a clear pattern for each field in terms of the number of book chapters per book.

*3.4 Effect sizes*

In table 5 we report the effect sizes between the groups analyzed for each of the three variables under study. We include Cohen's $d$ as well as correlation $r$ and use the convention proposed by Cohen to interpret the extent of the effect size, as mentioned earlier. As observed, for most groups, the effect size of the differences on their citation characteristics is small, with $d$ never reaching a value of 0.3. We only find medium effect sizes ($d \geq 0.30$) when analyzing differences by publisher type. In Engineering & Technology, we observe a medium effect size between Academic and Commercial Publishers ($d = 0.31$). In the areas of Arts & Humanities there is a medium effect size between Academic Publishers and University Presses ($d = -0.37$) and between Commercial Publishers and University Presses ($d = -0.43$). Finally, in Social Sciences we observe a medium effect size between Commercial Publishers and University Presses ($d = -0.38$).

**Table 5. Effect sizes between groups for each of the variables under study: presence of editors, inclusion in series and type of publisher.**

| Discipline | Type | d | r | Effect Size |
|---|---|---|---|---|
| ALL | Edited vs. Non Edited | 0.21 | 0.10 | Small |
| | Series vs. Non Series | 0.02 | 0.01 | Small |
| | Academic vs. Commercial | 0.25 | 0.12 | Small |
| | Academic vs. Univ Press | 0.03 | 0.02 | Small |
| | Commercial vs. Univ Press | -0.01 | -0.03 | Small |
| ENG | Edited vs. Non Edited | 0.28 | 0.14 | Small |
| | Series vs. Non Series | 0.09 | 0.05 | Small |
| | Academic vs. Commercial | 0.31 | 0.16 | Medium |
| | Academic vs. Univ Press | -0.15 | -0.08 | Small |
| | Commercial vs. Univ Press | -0.32 | -0.16 | Small |
| HUM | Edited vs. Non Edited | 0.12 | 0.06 | Small |
| | Series vs. Non Series | -0.06 | -0.03 | Small |
| | Academic vs. Commercial | 0.10 | 0.05 | Small |
| | Academic vs. Univ Press | -0.37 | -0.18 | Medium |
| | Commercial vs. Univ Press | -0.43 | -0.21 | Medium |
| SCI | Edited vs. Non Edited | 0.24 | 0.12 | Small |
| | Series vs. Non Series | 0.11 | 0.05 | Small |
| | Academic vs. Commercial | 0.09 | 0.04 | Small |
| | Academic vs. Univ Press | -0.08 | -0.04 | Small |
| | Commercial vs. Univ Press | -0.09 | -0.05 | Small |
| SOC | Edited vs. Non Edited | 0.16 | 0.08 | Small |
| | Series vs. Non Series | -0.02 | -0.01 | Small |
| | Academic vs. Commercial | 0.15 | 0.07 | Small |
| | Academic vs. Univ Press | -0.27 | -0.13 | Small |
| | Commercial vs. Univ Press | -0.38 | -0.19 | Medium |

*3.5 Citation characteristics of Highly Cited Books*

In Figure 1, we include the citation characteristics of the HCB for the four macro-areas under study according to the presence of editors, inclusion in series and publisher type. Of 1534 HCB, 65% were edited while 35% were non-edited. This general pattern also appears in three of these areas, especially in Science with 90% of HCB being edited books, followed by Engineering & Technology (65%) and Social Sciences (57%). The only exception is found in Arts & Humanities where the percentage of edited HCB is lower than that for non-edited with 42% of the total share of HCB. However, when interpreting the data for HCB presented in Figure 1, the total share of the BKCI (included in table 2) must be taken into account. For instance, in Engineering & Technology there is a higher share of edited books. In the case of the Arts & Humanities, edited books represent only 28% of the total share. However, the share for HCB edited is 42% of the total HCB in this area. This means that HCB are more commonly edited than non-edited books.

**Figure 1. Citation characteristics of the 5% most highly cited books in the Book Citation Index for the areas of Engineering & Technology, Arts & Humanities, Science and Social Sciences. 2005-2011 time period.**

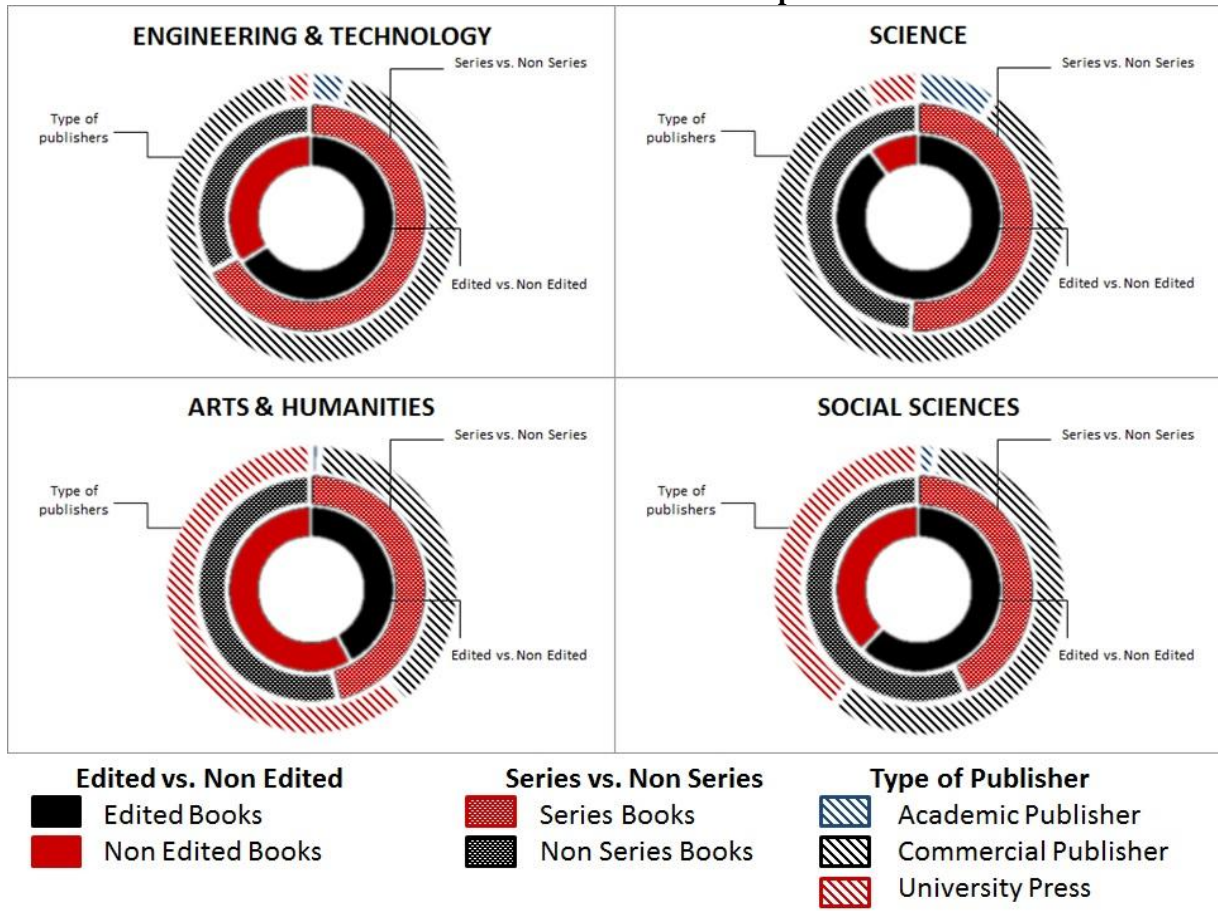

If we focus on HCB included in series, we observe that 65% of most cited books overall belonged to a series. Also, books included in series are better represented in the area of Engineering & Technology, while in Science the proportion is 50%. In Social Sciences and Arts & Humanities, the HCB with greater presence are those not included in series. In the four areas analyzed there are no significant differences in the overall distribution of books included and non-included in series as well as in the distribution of HCB. Therefore, while 71% of books in Engineering & Technology are included in series, 67% of the HCB in this area are also included in series, following a similar distribution.

Regarding the distribution of HCB according to the publisher type, the most significant result is the high representation of HCB among books from university presses, almost always higher than the other two publisher types for all areas. For instance, for all areas university presses represent 14% of the total share. However, when focusing on HCB, they represent 30%. This phenomenon is especially relevant in two of the four areas under study. Thus, in Arts & Humanities 28% of the total share are published by university presses, but 61% of the HCB are from this type of publisher. Similarly, in the Social Sciences they represent 15% of the total share but have 40% of the total HCB.

## 4. Discussion and concluding remarks

This paper analyzes the citation characteristics of books according to three variables: the presence of editors, their inclusion in series and the type of publisher. For this, we used a sample of 28634 books indexed in the Book Citation Index for four macro-areas during the 2005-2011 time period. We emphasize that the BKCI is a novel database constructed by Thomson Reuters which opens new opportunities for analyzing citation phenomena in books as it currently occurs with journals, where these characteristics have already been thoroughly analyzed (e.g., Peritz, 1981; Aksnes, 2003). By comparison with journal articles, many issues remain unsolved when analyzing books and their different typologies (i.e., edited, non-edited, included in series, etc.) for bibliometric purposes. Just as we know that review articles tend to receive more citations than research articles or that editorial material receives few citations, we still need to improve our understanding of the way the type of book affects their citation pattern. In this paper, we explore this issue through a large multidisciplinary dataset divided into broad research areas.

Our findings suggest that there are differences according to these factors. Despite acknowledging their importance, especially in the fields of the Social Science and Humanities, the bibliometric community has lacked data sources that would allow us to retrieve relevant citation and publication data. The emergence of Google Books or the Book Citation Index has made this possible. But, before embracing them and developing bibliometric indicators for research evaluation purposes just as we have with journal articles we must fully analyze and understand their citation characteristics in order to adjust and adapt them correctly to our bibliometric toolbox (Leydesdorff, 2009). This is important especially at a time when efforts are being redoubled in order to obtain bibliometric measures regarding the research evaluation of books. Despite finding differences when comparing types of book, the effect size of these differences is not as large as would be expected (table 5), this may be due to the coverage of the database which has been reported as having serious limitations (Torres-Salinas et. al, 2013b). In order to test the extent of these differences, further research is needed; focusing on specific fields. Furthermore, more thorough analysis of the conceptual differences between edited and non-edited books is needed. Should they receive a different treatment when being subjected to bibliometric analysis? Should we descend to the level of book chapter only when dealing with edited books?

Another issue to consider has to do with the consequences and possibility of applying publication and citation analysis to books for research evaluation purposes. In contrast to findings reported by Gorraiz, Purnell & Glänzel (2013), our data does include affiliation information which could be used to analyze research units such as countries or institutions. This difference may be due to the time gap between the downloading of the respective studies. Torres-Salinas et al. (2012) suggested developing rankings of publishers according to their citation data. In this study, the differences when larger effect sizes were found related to comparisons between publisher types. Perhaps this should be taken into account before attempting this goal. Table 6 shows the main findings of the study.

**Table 6. Highlights of the main findings of this study analyzing the factors which determine the citation of books in four major areas. Data: Book Citation Index. 2005-2011**

|  | **ENG** | **HUM** | **SCI** | **SOC** |
|---|---|---|---|---|
| **CHARACTERISTICS OF THE BKCI COVERAGE** | | | | |
| **Edited Vs Non Edited** | Edited and non-edited books are equally distributed | There are more non-edited books than edited (71%) | Edited and non-edited books are not equally distributed | There are more non-edited books than edited (60%) |
| **Series Vs Non Series** | Most books are included in series (71%) | Books included and not included in series are equally distributed | Most books are included in series (66%) | Books included and not included in series are not equally distributed |
| **Type of Publisher** | Most books are from commercial publishers (97%) | Most books are from commercial publishers (71%) | Most books are from commercial publishers (87%) | Most books are from commercial publishers (83%) |
| **CITATION CHARACTERISTICS OF BOOKS INCLUDED IN THE BKCI** | | | | |
| **Edited Vs Non Edited** | Edited books are more cited | Edited books are more cited | Edited books are more cited | Edited books are more cited |
| **Series Vs Non Series** | Books included in series are more cited | Books not included in series are more cited | Books included in series are more cited | There are no citation differences |
| **Type of Publisher** | Books from university presses are more cited | Books from university presses are more cited | Books from university presses are more cited | Books from university presses are more cited |

1) Non-edited books are more common in the Arts & Humanities and the Social Sciences than edited books. This could lead us to presume that non-edited books would be better more cited. However, edited books have a greater impact than non-edited books in all areas. This may be due to the effects of working collectively with a more diversified content and therefore, more chances of being cited. Another possible explanation may lie in the average number of book chapters per book. As observed in table 2, edited books have on average a much higher number of chapters; that is citable items, as conceived by the Book Citation Index. This may enhance their opportunity of receiving more citations than non-edited books.

2) The inclusion of books in series is more frequent in the areas of Engineering & Technology and Science, while in the Arts & Humanities and Social Sciences, their distribution is more homogeneous. However, the impact of books according to their inclusion in series varies depending on the area. In Engineering & Technology and Science, books included in series receive more citations than those which are not, although this is not as significant as in other cases. In the Arts & Humanities, books not included in series are those with higher impact, but there are fewer differences. In the case of the Social Sciences, the differences are almost negligible.

3) Considering publisher type, most books indexed in the BKCI belong to commercial publishers, especially in the areas of Engineering & Technology and Science. Though the distribution is similar in the Arts & Humanities and the Social Sciences, the university presses are better represented in these areas. However, one phenomenon is common across all areas: books published by university presses receive significantly more citations than the others. At this point we must approach this statement with caution as, after examining these publishers, we find that the university presses included in the BKCI are considered of huge prestige such as Cambridge UP, Princeton UP and University of California P. That is, books from university presses may be highly cited due to the better selection process of books and topics than that followed by commercial publishers, such as Elsevier, Routledge or Palgrave, for instance.

Finally, we must point out that the results offered in this analysis inherit the shortcomings of the database from which the data was retrieved. The BKCI is an on-going project which still shows significant limitations. Some of these may affect the results presented such as a bias towards English language publications (96% of its books are written in this language and 75% of the publishers come from the United Kingdom or the United States) and a great concentration of publishers. For example, Springer, Palgrave and Routledge alone account for 50% of the total database. Therefore, these issues must be taken into consideration when analyzing our findings. However, the large data set used may be a significant step towards a better comprehension of the citation characteristics of books. Also, this study presents a global overview of the scientific knowledge, dividing it into broad areas. In order to deepen on the citation characteristics of books further analyses are needed focusing on specific disciplines, especially from the fields of Social Sciences and Humanities.

## Acknowledgments

Thanks are due to the anonymous referee for their helpful comments. Nicolás Robinson-García is currently supported by a FPU grant from the Spanish Ministerio de Economía y Competitividad of the Spanish government. Bryan J. Robinson of the University of Granada Department of Translation and Interpreting revised the final version of the manuscript.